# Physical and Behavioral Adaptations to Prevent Overheating of the Living Wings of Butterflies

Cheng-Chia Tsai[1], Richard A. Childers[2], Norman Nan Shi[1]§, Crystal Ren[1], Julianne N. Pelaez[2]†, Gary D. Bernard[3], Naomi E. Pierce[2,4]*, and Nanfang Yu[1]*

[1]Department of Applied Physics and Applied Mathematics, Columbia University, New York, NY 10027, USA

[2]Department of Organismic and Evolutionary Biology, Harvard University, Cambridge, MA 02138, USA

[3]Department of Electrical Engineering, University of Washington, Seattle, WA 98195, USA

[4]Museum of Comparative Zoology, Harvard University, Cambridge, MA 02138, USA

*Correspondence to: N.Y. (email: ny2214@columbia.edu), N.E.P. (email: npierce@oeb.harvard.edu)

†Current address: Department of Integrative Biology, University of California, Berkeley, Berkeley, CA 94720, USA.

§Current address: Western Digital, San Jose, CA 95119, USA

## Abstract

The wings of Lepidoptera contain a matrix of living cells whose functioning requires appropriate temperatures. However, given their small thermal capacity, wings can overheat rapidly in the sun. Here we analyze butterfly wings across a wide range of simulated environmental conditions, and find that regions containing living cells are maintained at cooler temperatures. Diverse scale nanostructures and non-uniform cuticle thicknesses create a heterogeneous distribution of radiative cooling that selectively reduces the temperature of living structures such as wing veins and androconial organs. These living tissues are supplied by circulatory, neural and tracheal systems throughout the adult lifetime, indicating that the insect wing is a dynamic, living structure. Behavioral assays show that butterflies use wings to sense visible and infrared radiation, responding with specialized behaviors to prevent overheating of their wings. Our work highlights the physiological importance of wing temperature and how it is exquisitely regulated by structural and behavioral adaptations.

**Introduction**

The wings of Lepidoptera (butterflies and moths) are important organs both in flight and in signaling, where they have been studied intensively in the context of sexual selection[1-3], warning coloration[4,5], mimicry[6-8], and camouflage[9,10]. Although fully expanded wings are primarily composed of lifeless membranes covered by scales, mature wings actually contain complex networks of cellular structures, with, for instance, an extensive distribution of sensilla along wing veins[11-15]. Campaniform sensilla are used to sense the strain or deformation of the wings[14-18], and bristle sensilla found along wing veins and margins sense wing beat frequencies[12,19]. The afferent feedback from these mechanical sensilla to the central nervous system is critically involved in the generation of normal flight in insects[14-19]. Although detailed studies focus on only a few species of Lepidoptera, the taxonomic range of the species surveyed suggest that these are general features of lepidopteran wings.

Networks of living tissues typically require a constrained range of temperatures for optimal performance. Temperature regulation is critical in insects, with even minor changes in ambient temperatures having profound effects[20-23]. Much of the research on thermoregulation in adult Lepidoptera has focused on thoracic temperature and flight[24-35]. Little research has been done on the thermoregulatory and thermodynamic properties of the wings themselves. This is partly because of the difficulty of applying common temperature measurement techniques to such thin and light objects with small thermal capacity. In a pioneering early work, Wasserthal and Schmitz[36] implanted thermistors into wings and other parts of butterflies and then heated the insects with a calibrated light beam. This technique demonstrated that antennae, wings and thorax heat up at different rates and reach specific excess temperatures. However, these implants have the drawback that they can substantially change the local thermal capacity of the wing and lead to inaccurate measurements of both wing temperatures and the dynamics of temperature change. Additional research on the temperature of butterfly wings has relied upon detecting thermal radiation generated by the wings using infrared cameras[37]. However, butterfly wings are largely semitransparent in the mid-infrared spectrum (except for the thick veins near the base of the wing), meaning that infrared radiation detected by the camera is only partially contributed by the wings, leading to errors in estimating the temperature over the entire wing.

In this work, we investigate the thermodynamic and thermoregulatory properties of the wing itself, first by observing the distribution and persistence over the entire adult lifespan of living tissues in the wings of butterflies, and then by studying the thermodynamics of the wing and the physical and behavioral adaptations that modulate wing temperatures in ways that protect these tissues. Using a robust ecological model species, *Vanessa cardui*[38,39], we show that circulatory and tracheal systems remain active in the wing veins throughout the entire adult stage. We develop an infrared hyperspectral imaging technique that permits accurate temperature mapping of objects that are both lightweight and translucent in the mid-infrared. We find that wings are extremely heterogeneous in their thermodynamic properties, with wing veins and other living parts of the wing being cooler than inter-vein regions under solar radiation. This is achieved via elevated thermal emissivity due to a thickened chitinous layer and specialized nanostructuring of their scales such that they more efficiently dissipate heat through thermal radiation.

In addition to these structural modifications, butterflies employ behavioral strategies to control the heating of their wings by the external environment. We present the responses of species of butterflies representing 6 families with different sizes, shapes and coloration to thermal stimulations of the wings. Even heating by a small localized laser spot is sufficient to elicit displacement behaviors, and there is a tight window of trigger temperatures across families.

Butterflies bask for warmth at cool temperatures, and here again, previous studies have focused on thoracic temperature thresholds necessary for flight[24-35]. We find that when butterflies settle and convective cooling of the wings is substantially reduced, the wings can overheat quickly (within 10 seconds) under strong solar radiation, whereas other bulkier body parts (head, thorax, abdomen), whose temperatures are less volatile, remain in a thermally comfortable window. It is therefore the wing temperatures, rather than the thoracic ones, that determine when basking ceases. These physical and behavioral adaptations underscore the importance of wings as highly responsive living structures and the centrality of wing thermoregulation to lepidopteran biology.

**Results**

**Butterfly wings as living organs**

We conducted morphological studies of the wing micro-structures after carefully removing scales from both sides of the wings of living *Vanessa cardui*, *Satyrium caryaevorus*, and *Parrhasius m-album* butterflies. We selected these species both for pragmatic and biological reasons. *Vanessa cardui* (**Fig. 1a**) is a well-known ecological organism because it occurs worldwide, engages in long-distance migration, is highly polyphagous, and can be purchased and raised on artificial diet. The other two species were chosen because they possess androconial scent pads and are locally available and easy to collect and observe in the field.

Campaniform and bristle sensilla similar to those found in other lepidopteran species[12-15] are present in all the wing veins we have inspected using optical microscopy, including interior and marginal veins, and each vein contains multiple sensilla (**Fig. 1b,c**, **Supplementary Fig. 1d**). Staining using methylene blue[40-42] shows that each sensillum is connected via axons to a network of nerves distributed throughout the wing veins (**Fig. 1c-e**, **Supplementary Figs. 1** and **2**, **Supplementary Movies 1** and **2**).

The functional morphology of wings has been described in a number of insect orders[43-46], but the relationship between the circulatory and tracheal systems in wings has only been explored in detail for a few groups[11]. In *V. cardui*, we observed that wing veins provide pathways for the tidal flow of air and hemolymph (**Fig. 1f,g**, **Supplementary Movie 3**). Each of the larger veins contains a central tracheal tube and one or two hemolymph channels (**Fig. 1f**, **Supplementary Fig. 4b**). Large, semitransparent cells (i.e., hemocytes) can be observed flowing inside the channels, and enable continuous monitoring of the hemolymph flow direction and speed. The flow of hemolymph changes directions rhythmically. Given that the tracheal tube and the hemolymph channels occupy the rigid wing vein, an increase in volume of one will concomitantly decrease the volume of the other: when the tracheal tube expands, air enters the wing vein, and hemolymph is drawn from the tip to the base of the wing and eventually into the thorax; when the tracheal tube decreases in diameter, air flows out of the wing vein, and hemolymph flows back out to the wing tips (**Supplementary Movie 3**). The frequency of this tidal flow of hemolymph and air in and out of the wing veins is on the order of once every few minutes when the butterflies are not agitated (**Fig. 1g**).

In *V. cardui*, hemolymph flow sustains over the lifespan of the adult. We quantified the hemolymph flow by the dilation and contraction rates of hemolymph channel width as a function of time, taking advantage of the strong correlation between hemolymph flow rate and the change in width of the hemolymph channel (**Supplementary Fig. 4c**). Dilation and contraction of hemolymph channels remain active for more than three weeks after eclosion, and the hemolymph flow does not weaken appreciably with age (**Fig. 1g**, $n = 22$ individuals tested).

Males of a number of species of Lepidoptera have androconial organs on their wings involved in the generation and dissemination of pheromones[47-51]. These alar androconial organs are extremely diverse, and we focused our analysis on only two kinds of androconial organs of the hyperdiverse tribe Eumaeini (Lepidoptera: Lycaenidae), termed 'scent pads' and 'scent patches'[52,53]. Scent pads are unique to the Eumaeini, and occur in small regions of the wing where the wing membranes are not fused, thereby forming intermembrane pockets; scent patches are found in many other groups of butterflies as well as in the Eumaeini, and occur in regions of the wing where the dorsal and ventral membranes are fused. Both scent pads and patches are covered by modified scales that are distinct from scales on other parts of the wing. They can exist alone or together on a wing. When found together, they are almost always adjacent to each other.

We observed that scent pads of *S. caryaevorus* and *P. m-album* butterflies both contain intricate networks of hemolymph channels and living tissues (**Fig. 2b,d**, **Supplementary Fig. 2d**). We found that the flow of hemolymph through the scent pads is rhythmical and unidirectional: it enters the organ from a costal vein defining the leading edge of the discal cell, spreads over the entire intermembrane pocket, and drains into the anterior vein of the discal cell (**Fig. 2b,d,e**, **Supplementary Movies 4-7**). This is different from the tidal flow pattern observed in the veins of *V. cardui*. In both *S. caryaevorus* and *P. m-album*, we observed a "wing heart" that beats endogenously at a rate of a few dozen times per minute and appears to draw hemolymph unidirectionally through the scent pad (**Fig. 2b,d**, **Supplementary Movies 4** and **5**).

**Whole-spectrum response of butterfly wings**

Extensive studies have established that the visible colors of butterfly wings are created by the deposition of pigments in the wing scales and/or optical diffraction by nanostructuring of the scales[54-59]. Here we examined butterfly wings not only under light that is visible to the human eye, but also under wavelengths beyond the visible, since the whole-spectrum response of butterfly wings, from the ultraviolet to the visible to the mid-infrared, governs their thermodynamic properties[60-63] (**Fig. 3**).

Consider a male *Bistonina biston* (Lycaenidae: Eumaeini) (**Fig. 4a**). The dorsal forewing surface is covered by at least four types of wing scales (**Fig. 4d**), and the wing contains androconial organs consisting of a 'scent patch' and a 'scent pad', both covered by modified scales. Spectroscopic measurements show that, with dorsal irradiation, the region covered by black scales has stronger solar absorption (i.e., absorptivity integrated over the solar spectrum of 0.71) than the lighter-colored scent patch, scent pad, and structural blue region that have integrated solar absorptivities of 0.60, 0.63, and 0.59, respectively (**Fig. 4b** upper panel). The lighter-colored ventral wing surface (**Fig. 4b** lower panel) has reduced solar absorption: the four ventral regions that correspond to the black region, scent patch, scent pad, and blue region on the dorsal side have integrated solar absorptivities of 0.57, 0.59, 0.60, and 0.45, respectively. All wing regions have substantially lower solar absorptivity in the near-infrared ($\lambda = 0.7-1.7$ μm) than in the visible (**Fig. 4b**); this helps to reduce the temperature of the wings, which are more susceptible to overheating by the sun than bulkier body parts.

While the visible wing color is related to the absorption and reflection of sunlight, it has little correlation with the wing's ability to cool down via thermal radiation. Those parts of the heated wing with higher emissivity are better able to dissipate heat through thermal radiation to a cool environment (e.g., the sky and terrestrial environment)[60,62,64] (**Fig. 3**). We conducted hyperspectral imaging of the wings of an adult *B. biston* in the mid-infrared ($\lambda = 2.5–17$ μm) and found that different parts of the butterfly wing have dramatically different thermal emissivities. In particular, the scent patch, scent pad, and wing veins have emissivities approaching unity (i.e., the emissivity of a perfect blackbody) (**Fig. 4c,e**). Our spectroscopic studies of individual wing scales show that the scent patch scales are by far the most emissive among all types of wing scales on *B. biston* (**Fig. 4g**). We found that the high thermal emissivity of the scent patch is caused by unique nanostructures of the scales (**Fig. 4d,h,i**), whereas the high emissivities of the scent pad and the wing veins are due primarily to the physical thickness of their unfused membranes (**Fig. 4f**). The scent patch scales of other species of Eumaeini are similarly nanostructured as those found on the wings of *B. biston,* and also possess elevated emissivity (**Supplementary Fig. 9**). We have characterized the infrared properties of the wings of individual adults representing more than 50 species (**Supplementary Tables 6** and **7**) and found that scent patches, scent pads, and wing veins in all studied species, without exception, have elevated thermal emissivity and thus enhanced radiative-cooling capabilities (**Supplementary Fig. 18-31**).

**Wing temperatures under simulated environmental conditions**

We developed a noninvasive technique based on infrared hyperspectral imaging to accurately determine wing temperatures under experimental conditions that simulate the butterflies' natural environment (**Supplementary Fig. 33**). By measuring mid-infrared emissivity, transmissivity, and reflectivity spectra of the butterfly wing, we can distinguish thermal radiation generated directly by the wing from environmental thermal radiation that reflects from and transmits through the wing (**Supplementary Fig. 33c**), and thus derive temperature distributions across the entire wing (**Supplementary Fig. 36**).

Temperature maps for the forewings of six Eumaeini butterflies are shown in **Fig. 5**. We observed that, despite the wings' wide variation in visible coloration and pattern, the average temperature of the scent patches, pads, and wing veins that contain living cells (including hemolymph cells, androconial cells, and mechanical and thermal sensilla) is always lower than that of the remaining "non-living" parts of the wings (i.e., regions with fused membranes). We also observed that a cooler radiative background (e.g., $T_c = -40^{\circ}C$) leads to reduced overall wing temperatures (**Fig. 5**, **Supplementary Fig. 38**) because of a larger net radiative heat transfer from the heated wings to the background. Under the condition of ventral irradiation of light equivalent to the full sun, the scent patches, pads, veins, and margins rarely exceed 45°C. This offers a plausible explanation for why many Eumaeini butterflies tend to close their wings when resting or basking, and infrequently expose their dorsal wing surfaces.

Bright visible and near-infrared coloration does help reduce solar absorption and thus reduce temperatures (e.g., the light gray ventral side of *Strephonota tephraeus*, **Fig. 5c**, and the white bands on the ventral side of *Panthiades aeolus*, **Fig. 5f**), but the largest temperature reduction is achieved by a combination of low solar absorption and efficient radiative heat dissipation: the white scent pad of *Rekoa meton* is ~15°C cooler than the hottest part of the wing (**Fig. 5a**), and the light-colored scent patch and pad of *B. biston* are ~10°C cooler than the hottest part of the wing (**Fig. 5b**). The cryptic scent patch on *Theritas hemon* becomes visible in the temperature maps due to its higher thermal emissivity and cooler temperatures (**Fig. 5d**).

The importance of high thermal emissivity in reducing wing temperatures is further exemplified in the case of dorsal irradiation of *T. mavors* and *P. aeolus*: the temperature of their black scent patches and pads, which have high solar absorptivity but also near-unity thermal emissivity, is 2 to 5°C cooler than the hottest part of the wing (**Fig. 5e,f**).

**Wings as temperature sensors that modulate behavior**

The importance of wing temperatures can also be demonstrated by behavioral experiments. We conducted the study on species representing 6 of the 7 recognized families of butterflies (all but Hedylidae) that were readily available to us. During each experiment, a collimated laser beam (either in the visible or near-infrared spectrum) was directed onto a butterfly wing with normal incidence to increase the wing temperature locally. Laser intensity was selected such that the temperature increased from 25°C to 45°C in 7–8 seconds. Butterflies were observed to move in characteristic ways to displace thermal stimuli applied to their wings (**Fig. 6a-g**). These reactions could be reliably elicited by positioning the laser spot anywhere along the wing veins, or on the androconial organs, indicating that the "living" components in the wings contain a distributed network of thermal sensors (**Supplementary Fig. 44**, **Supplementary Movies 8** and **9**).

We used thermal camera videos to record the peak wing temperature, $T_{trigger}$, at which butterflies started to show displacement responses. We observed that butterflies across diverse families converge on a $T_{trigger}$ of about 42°C ($n$ = 50 species, $T_{trigger}$ = 41.89±1.54°C) (**Fig. 6h**). Butterflies in the family Papilionidae show the highest family average $T_{trigger}$ of 43.55±1.53°C; butterflies in the Hesperiidae and Lycaenidae show low average $T_{trigger}$ of 41.15±1.04°C and 40.78±1.00°C, respectively. Further research on wing solar absorptivity and thermal emissivity in the context of the ecology of each species would be necessary to explain the subtle differences in $T_{trigger}$. In 10 out of 11 species with sufficient sample sizes for comparison, females have a higher median $T_{trigger}$ than the males (**Fig. 6h**). The exception to this pattern among the butterflies we analyzed is the Zabulon skipper, *P. zabulon*, where females have lower $T_{trigger}$ than males. These butterflies are sexually dimorphic, and the dark brown females may overheat more readily than light-colored males.

We also studied the basking behavior of butterflies under experimental conditions simulating natural environmental conditions (**Supplementary Movies 10** and **11**). Adults of three species of hairstreak, *P. m-album, S. favonius,* and *S. caryaevorus* exhibit lateral basking[24,25] behavior by closing their wings dorsally and tilting sideways to present their ventral wing surfaces at right angles to the light source (**Fig. 7a-c**). In laboratory experiments, we used a Xenon lamp shining uniformly onto a substrate from above to simulate sunlight. We observed that the percentage of time the butterflies spent basking decreased as the ambient temperature, $T_a$, increased (**Fig. 7e**). Under the same experimental conditions, adults of *S. caryaevorus* basked far less than did those of *P. m-album* and *S. favonius*. Additional work would be needed to explain differences in basking behaviors, but differences in heating caused by coloration could explain the observed variation in basking time between these species.

We monitored the thoracic and wing temperatures of individuals of *S. caryaevorus* under experimental conditions (ambient temperature of 21.5°C and light intensity of 66.0 mW/cm², corresponding to 2/3 of full sun in the field) and found that the effect of basking on heating up the wings was considerably stronger and faster than its effect on heating up the thorax because of the relatively small thermal capacity of the wings (**Fig. 7f**). For example, basking increased the temperature of the wing veins by 4.48±2.25°C, reaching 39.5±1.52°C, and increased the temperature of the scent pads by 6.60±2.19°C, reaching 40.0±1.54°C, while the temperature of the thorax increased by only 1.01±0.54°C during the basking event. The data suggest that cessation of basking occurs when the wing temperatures are too high (i.e., ~40°C) whereas thoracic temperature remains relatively constant during the basking. This thoracic homeostasis may allow the butterflies to remain flight-ready.

Similar experiments to monitor the thoracic and wing temperatures of the two hairstreaks, *S. favonius* ($n = 4$) and *S. caryaevorus* ($n = 6$) when basking under a lamp corresponding to 2/3 of full sun and at a variety of ambient temperatures showed that the peak wing temperature at the moment when the butterflies terminated basking was 4.08°C higher than that of the thorax, for *S. favonius* (**Fig. 7g**), and 8.04°C higher than that of the thorax, for *S. caryaevorus* (**Fig. 7h**), again supporting the hypothesis that butterflies stop basking when wing temperatures have become too high.

In our field studies using a thermal camera, we found that *S. caryaevorus* butterflies almost always bask in the sun within a few seconds after ceasing flight (**Supplementary Fig. 45**, **Supplementary Movie 12**). At that moment, the thoracic temperature is reasonably high (>30°C) as a result of the metabolic heat generated by the flight muscles, but the wings, due to strong convection during flight, have temperatures close to the ambient temperature. Our lab experiments confirmed that at the moment when the *S. caryaevorus* butterflies stopped flight, their thoracic temperatures were in the range of 30.9°C±0.73°C, while their wing temperatures were only 0.73±0.47°C above the ambient temperature, which was varied from 25°C to 28°C (**Fig. 7i**). Basking behavior may thus be elicited by cold wings: presumably the temperature of the wings must be sufficiently high for optimal functioning of mechanical sensilla, which are crucial for such agile flying insects.

**Discussion**

While previous work on lepidopteran wings has primarily focused on visible patterns involved in inter- and intra-specific signaling, this work emphasizes the interactions of the wings as physical structures with the abiotic electromagnetic environment, including solar and thermal radiation. Our results show that butterfly wings play a largely unappreciated role in sensing solar radiation, and that their spectral properties across wavelengths beyond the visible, including the infrared solar and thermal spectra, are tailored to help them better adapt to the electromagnetic radiation environment. Selection has favored adaptations such as heterogeneous cuticle thickness, specialized scale nanostructures, contrasting dorsal and ventral wing coloration and finely-tuned behavioral reactions to protect the vulnerable living tissues within wings.

When butterflies close their wings, exposing only their ventral sides, the thermal emissivity of the closed wings is larger than that of a single wing due to their increased physical thickness. In the Eumaeini (Lycaenidae), such an increase of thermal emissivity is large for the non-living regions with fused membranes (e.g., from 0.30 to 0.51), moderate for the wing veins (e.g., from 0.70 to 0.91), and small for the scent patches and pads (e.g., from 0.95 to 0.998). Thus, closing the wings would reduce primarily the temperature of the regions with fused membranes, and because wing veins, scent pads and patches are bounded by regions with fused membranes, their temperatures would also decrease via thermal conduction.

An additional effect of wing closure is that the time constant of temperature changes increases substantially because of the increased thermal capacity. Thus dorsal basking with open wings and lateral basking with closed wings could have different effects on the thermodynamics of butterfly wings. In tropical and temperate climates, lateral basking with enhanced thermal emissivity and increased thermal capacity of the stacked wing pairs enables butterflies to bask longer to warm up their thoraces while reducing the risk of overheating the wings; in contrast, butterflies at high latitudes or altitudes could warm up their wings efficiently by dorsal basking in the sunlight with their wings fully or partially open.

In *The Naturalist on the River Amazons*[65], Henry Walter Bates writes about butterfly wings, "On these expanded membranes Nature writes, as on a tablet, the story of the modifications of species, so truly do all changes of the organization register themselves thereon…" This study offers a perspective on butterfly wings as living organs through an analysis of their thermodynamic and thermoregulatory properties. We have considered the synergistic effects that the electromagnetic environment has had in shaping the morphology, physiology and behavior of butterflies, and identified physical properties and behaviors that are likely to be applicable to the flying wings of most insects when exposed to sunlight.

## METHODS

### Observing micro-structures and hemolymph flow in wings

*Vanessa cardui*, *Satyrium caryaevorus*, and *Parrhasius m-album* butterflies were anesthetized with $CO_2$ and scales were carefully removed from both sides of the wings with a paintbrush. Each specimen was then secured using a low melting temperature wax to two microscope slides, one on either side of the body (**Supplementary Fig. 4a**). Photos and videos were taken with a microscope equipped with a CCD camera. The direction and speed of hemolymph flow (**Figs. 1g and 2b,d,e**, **Supplementary Fig. 5c**) were inferred from the movement of semitransparent cells (hemocytes) inside the hemolymph channels and scent pads (**Supplementary Movies 3-7**). Stacking photography was sometimes used to better resolve fine structures within the wings (**Figs. 1b-e** and **2b,d**, **Supplementary Figs. 1-3** and **5**).

Methylene blue, a dye that has been widely used to stain nervous tissues[40,42] including the neurons of Lepidoptera[41], was used to stain the neurons in wing veins. Adult *V. cardui* and *S. caryaevorus* butterflies were anesthetized. A saturated aqueous solution of methylene blue (Aldon Corporation) was injected into the thoraces: 50-100 µl used for individual *V. cardui* and 10 µl used for individual *S. caryaevorus*. The stained wings were removed 6 hours after injection, when the butterflies had recovered from anesthesia and the injected stain had been circulated by the hemolymph into the wing veins. The scales were removed from both sides of the stained wings and stacking photography was used to take pictures of stained wing tissues.

### Infrared hyperspectral imaging of butterfly wings

The infrared optical properties of butterfly wings were measured using a hyperspectral imaging technique. To measure the transmissivity distribution, the wing was secured by two pairs of thin threads and then raster scanned under an infrared microscope (Bruker Hyperion 2000), with one transmission spectrum taken at each position. The raster scan generated a three-dimensional ($x,y,\lambda$) optical transmission hyperspectral data cube, where $\lambda$ includes 1763 wavelengths ranging from 2.5 μm to 16.7 μm and the step of scan is $\Delta x = \Delta y = 90$ μm (comparable to the sizes of individual wing scales). Similarly, an optical reflection hyperspectral data cube was generated by collecting spectra of reflected infrared light at all ($x,y$) positions over the butterfly wing. An emissivity hyperspectral data cube was then created by calculating the absorption spectrum at each position using 1 − transmissivity − reflectivity, as Kirchhoff's law of thermal radiation states that radiative emission and absorption are reciprocal processes, and thus emissivity equals absorptivity[64]. The hyperspectral data cubes (simplified version of the data cube for *B. biston* shown in **Fig. 4e**, **Supplementary Fig. 36b**, those of other species in **Supplementary Figs. 18–31**) enable us to visualize butterfly wings at different infrared wavelengths. Features that are not discernible to the naked eye stand out, and these features have little correlation with visible patterns of butterfly wings. The hyperspectral data cubes were used to derive the temperature distribution on the butterfly wings.

**Infrared properties of chitin used in full-wave simulations**

The membranes and scales of butterfly wings are mainly composed of the long-chain polymer chitin $(C_8H_{13}O_5N)_n$ which is the primary component of the exoskeletons of arthropods, including some crustaceans (such as crabs and shrimps) and insects[66]. The optical properties of chitin have not been studied thoroughly due to the difficulty in the production of optical-grade materials and its various derivatives depending on the sources. An average refractive index of 1.56 has been used in many publications and previous studies reported the dispersive complex refractive indices of shrimp chitin samples only in the UV and visible spectral range ($\lambda = 250–750$ nm)[67,68]. Here, the dispersive complex refractive indices of chitin composing the butterfly wings in the mid-infrared range ($\lambda$ = 2.5–20 μm) were derived from measured spectra and applied in full-wave simulations to investigate the interactions between mid-infrared light and nanostructured wing scales.

The chemical composition and molecular structure of chitin have been identified using various methods including infrared spectroscopy and X-ray diffraction[69,70]. Thirteen resonance frequencies in the mid-infrared spectral range corresponding to the major vibrational and rotational bands of chitin molecules have been characterized (**Supplementary Table 2**) and applied in the Lorentz oscillator model[71],

$$\frac{\varepsilon(\omega)}{\varepsilon_0} = 1 + \sum_m [\varepsilon_r(\omega_{0,m}, \omega_{p,m}, \gamma_m) + i\varepsilon_i(\omega_{0,m}, \omega_{p,m}, \gamma_m)]$$
$$= 1 + \sum_m \left[\frac{\omega_{p,m}^2(\omega_{0,m}^2 - \omega^2)}{(\omega_{0,m}^2 - \omega^2)^2 + \omega^2\gamma_m^2} + i\frac{\omega_{p,m}^2\gamma_m\omega}{(\omega_{0,m}^2 - \omega^2)^2 + \omega^2\gamma_m^2}\right], \quad (1)$$

to calculate the complex refractive indices $n_c(\omega) = n(\omega) + ik(\omega) = \sqrt{\varepsilon(\omega)}$, which were then used in the transfer matrix method to calculate the absorption spectrum of a thin film of chitin. In Equation (1), $m$ labels the various vibrational and rotational bands of chitin, $\varepsilon_o = 8.854 \times 10^{-12}$ $C^2N^{-1}m^{-2}$ is the permittivity of free space, and $\omega_{o,m}$ is the $m^{th}$ resonance frequency. The values of plasma frequency, $\omega_{p,m}$, and collision frequency, $\gamma_m$, were simultaneously tuned to obtain the best fit between the calculated absorption spectrum and a measured spectrum from a piece of butterfly wing membrane with scales carefully removed, which can be considered as a relatively flat slab with a uniform thickness. The absorption spectrum of the scale-less wing membrane was measured using a Fourier transform infrared spectrometer and its thickness was determined using a scanning electron microscope. The obtained complex refractive indices of chitin were used in full-wave simulations to investigate the interactions between infrared light and micro-structures of wing scales (**Fig. 4h,i**).

**Experimental setup simulating natural radiative environment**

A butterfly wing absorbs sunlight and exchanges energy via thermal radiation with the ambient environment, including the terrestrial environment beneath the wing and the sky above (**Fig. 3**). The radiative energy loss from the heated wing to the cold sky represents a major channel of heat dissipation in addition to heat losses via convection.

In the thermodynamic experimental setup (**Supplementary Fig. 33**), the light produced by a Xenon lamp shining from above at a distance of about half a meter was used to simulate sunlight. The lamp occupied only a small solid angle in the top hemisphere. The lab space over the top hemisphere was maintained at $T_a = 25°C$ to simulate the terrestrial environment. A temperature controllable cryostat equipped with an infrared-transparent window underneath the wing specimen was used to simulate the sky in different weather conditions.

The cryostat consists of a vacuum cryostat chamber, a cooled plate located at the center of the chamber, and a potassium bromide (KBr) window. The cooled plate is in contact with a liquid-nitrogen-cooled cold finger and its temperature can be controlled within 0.01°C of a set temperature by using a Lake Shore cryogenic temperature controller. The top surface of the cooled plate that faces the window is coated with a thick layer of paint and has a near-unity mid-infrared emissivity. The KBr window, 1.5 inches in diameter, has a transmissivity >90% over a broad mid-infrared spectral range ($\lambda = 2.5$-$20$ μm). The negligible reflectivity of the window minimizes reflections within the window and between the window and the wing specimen. The inner surface of the cryostat chamber, which is at ambient temperature, is an aluminum oxide layer and has thermal emissivity approaching unity. The combination of the KBr window and the cooled plate can be treated as a planar substrate with a finite size, a controlled temperature, and a thermal emissivity of ~1. Wing specimens are placed outside of the cryostat chamber a few millimeters above the KBr window in still air (i.e., free convection) at the ambient temperature. The solid angle subtended by the cooled plate with respect to the wing specimen has a half angle of $\theta_c = 25.3°$.

Calculations indicate that the total amount of thermal radiation generated by the cryostat setup ranges from 289 to 415 W/cm², when the temperature of the cooled plate is tuning from $T_c$ = −40 to +20°C (**Supplementary Fig. 35c**). This range of thermal radiance matches with field data measured using a pyrgeometer[72] and with downward atmospheric thermal radiance calculated using theoretical models (i.e., the Swinbank model[73] and the RRTM model[74]) (**Supplementary Fig. 35a,b**), and thus effectively simulates the wide variety of environmental conditions that butterflies experience.

**Deriving wing temperatures from thermal camera images**

A noninvasive technique based on the results of infrared hyperspectral imaging was developed to map temperature distributions on butterfly wings. The infrared flux detected by the thermal camera (FLIR SC660) (examples for *B. biston* shown in **Supplementary Fig. 36a**) has three contributions (**Supplementary Fig. 33c**):

$$\Phi_{cam} = \Phi_w + \Phi_t + \Phi_r, \qquad (2)$$

where $\Phi_w$ is the thermal radiation produced by the wing, $\Phi_t$ is the thermal radiation produced by the cryostat and transmitted through the wing, and $\Phi_r$ is the thermal radiation produced by the surrounding environment and reflected by the wing:

$$\Phi_w(T_w) = 2\pi \int_0^{\theta_{lens}} d\theta \sin\theta \cos\theta \int_{\lambda_{cam}} d\lambda \left[ \varepsilon_w(\lambda, \theta) \frac{2hc^2}{\lambda^5} \frac{1}{exp(hc/\lambda k_B T_w) - 1} \right], \qquad (3)$$

$$\Phi_t(T_c) = 2\pi \int_0^{\theta_{lens}} d\theta \sin\theta \cos\theta \int_{\lambda_{cam}} d\lambda \left[ t_w(\lambda, \theta) \frac{2hc^2}{\lambda^5} \frac{1}{exp(hc/\lambda k_B T_c) - 1} \right], \qquad (4)$$

$$\Phi_r(T_a) = 2\pi \int_0^{\theta_{lens}} d\theta \sin\theta \cos\theta \int_{\lambda_{cam}} d\lambda \left[ r_w(\lambda, \theta) \frac{2hc^2}{\lambda^5} \frac{1}{\exp(hc/\lambda k_B T_a) - 1} \right]. \quad (5)$$

The sum of these three terms equals the radiance received by the camera:

$$\Phi_{cam}(T_a) = 2\pi \int_0^{\theta_{lens}} d\theta \sin\theta \cos\theta \int_{\lambda_{cam}} d\lambda \left[ \frac{2hc^2}{\lambda^5} \frac{1}{\exp(hc/\lambda k_B T_{cam}) - 1} \right]. \quad (6)$$

Here $\varepsilon_w(\lambda,\theta)$, $t_w(\lambda,\theta)$, and $r_w(\lambda,\theta)$ are, respectively, angle and wavelength dependent emissivity, transmissivity, and reflectivity; $\theta_{lens} = 12.5°$ represents the half viewing angle of the thermal camera lens; $h$ is the Planck constant; $c$ is the vacuum speed of light; $k_B$ is the Boltzmann constant; $T_w$ is the temperature of the wing to be determined, $T_c$ is the cryostat temperature, $T_a$ is the ambient temperature, and $T_{cam}$ is the temperature directly read from the thermal camera; the integration is carried over the wavelength range from $\lambda_{cam}$ =7 µm to 14 µm, which is the transmission window of the internal filter of the thermal camera. Note that the $cos\theta$ term indicates that emission of thermal radiation follows Lambert's cosine law.

A case study of the forewing of a *Bistonina biston* butterfly has shown that the angle dependence of $\varepsilon_w(\lambda,\theta)$, $t_w(\lambda,\theta)$, and $r_w(\lambda,\theta)$ is weak within the viewing angle of the thermal camera lens (i.e., 25 °) (**Supplementary Fig. 37b**); therefore, the values of $\varepsilon_w(\lambda,\theta)$, $t_w(\lambda,\theta)$, and $r_w(\lambda,\theta)$ measured at surface normal direction were used in the above equations to determine the wing temperature, $T_w$. This simplification only leads to an overestimation of the wing temperature by less than 1.2°C (**Supplementary Table 3**). With the hyperspectral data of the butterfly wing (i.e., spatial distribution of $\varepsilon_w(\lambda)$, $t_w(\lambda)$, and $r_w(\lambda)$ over the entire wing area), the temperature distribution over the whole butterfly wing can be determined (**Fig. 5, Supplementary Fig. 36**).

It is important to note that the second and third terms in Equation (2) do not contain information about butterfly wing temperatures, but can still make substantial contributions to the detected signal by the thermal camera. In particular, $\Phi_t$ is proportional to the transmissivity of the wing, $t_w$, which can have large values for the parts of the wing that are physically thin and not covered by specialized scales like those found on the scent patches. For example, $t_w$ is ~0.8 around $\lambda = 11$ μm and $t_w$ averaged over the entire spectrum window of the thermal camera (i.e., $\lambda = 7-14$ μm) is ~0.7 for the membranous (non-vein) blue and black regions of the *B. bison* forewing. In other words, butterfly wings can be fairly transparent in the thermal radiation spectrum, and direct temperature readings from the thermal camera, $T_{cam}$, can be highly misleading and inaccurate. For example, what the thermal camera detects from the membranous blue and black regions of the *B. bison* forewing has comparable contributions from $\Phi_t$ and $\Phi_w$; thus, it would report the effective temperature of a superposition of the warm wing and the cold cryostat behind the wing: $T_{cam}$ would be somewhere between $T_c$ and $T_w$, and considerably smaller than $T_w$ in the case of $T_c = -40^\circ$C (mimicking clear sky with low humidity). However, $T_{cam}$ of the regions of the wing with high thermal emissivity (i.e., veins, scent patches and pads) is much closer to the actual wing temperature, because these regions have low infrared transmissivity (transmissivity $\cong 1 -$ absorptivity $= 1 -$ emissivity) and thus the contribution of the cryostat behind the wing to the detected radiance by the thermal camera is small.

While $\Phi_t$ can be comparable to $\Phi_w$, $\Phi_r$ is typically one order of magnitude smaller than either $\Phi_t$ or $\Phi_w$, because $\Phi_r$ is proportional to the infrared reflectivity of the wing, $r_w$, which is only a few percent over the entire wing (**Supplementary Fig. 36b**). Nevertheless, including $\Phi_r$ into Equation (2) enables more accurate determination of $T_w$.

**Displacement in response to local heating of butterfly wing**

In the experiment (**Fig. 6**), an intact adult butterfly was placed in an arena where it could walk freely. The lid of the container was composed of a thin polyethylene membrane that is transparent to mid-infrared radiation (to facilitate observation using a thermal camera). Once the butterfly was standing still in the container while its wings and body reached thermal equilibrium with the ambient environment, a collimated laser beam (either a telecom laser at $\lambda = 1.55$ μm or a supercontinuum laser with broadband emission at $\lambda = 0.4–1.7$ μm) was then directed onto either its left or right wing (haphazardly chosen) with normal incidence to increase the wing temperature locally. The beam diameter was adjusted to cover a segment of a wing vein 2–5 mm in length, and the beam was typically pointed at the junction of the radial veins of the forewing.

As the temperature of the region illuminated by the laser beam increased to a threshold value, the butterfly would move, and this response was scored as a displacement event (**Supplementary Movies 8** and **9**). The temperature at which this occurred was recorded as the threshold temperature, $T_{trigger}$. The laser power was controlled such that the temperature increased from 25°C to 45°C in 7–8 seconds. The rate at which the temperature increased was chosen to prevent over-stimulation that could either startle butterflies into flight or cause irreversible damage to the sensory neurons in their wings.

A thermal camera was used to record videos of the response, from which $T_{trigger}$ was determined. Wing veins have thermal emissivity close to unity, so direct readings from the thermal camera were used as the accurate temperature of the wing vein. Similarly, we used the thermal camera to record temperatures during basking both in the field and lab, again taking advantage of the fact that both the thorax and the wing veins have thermal emissivity approaching unity.

Different butterfly species move their wings in characteristic ways to displace thermal stimuli applied to the wings. For example, butterflies in the family Papilionidae with wings opened initially at rest tend to flap their wings, and butterflies in the family Lycaenidae with wings closed initially at rest tend to walk forward or turn around to avoid the laser spot (**Supplementary Movie 8**). Once the movement of the butterfly was observed, we blocked the laser beam immediately to prevent further increase of the wing temperature and possible damage caused by overheating.

Care was taken to ensure that the response was solely due to the temperature of the illuminated wing region exceeding a threshold value. Specifically, the laser spot was never positioned on wing regions close to the base of the wing, which is also close to the insect's eye. Movement and lighting changes in the surrounding environment were minimized to prevent accidentally stimulating the butterfly during the experiment. Control experiments were conducted that indicated butterflies may not be able to see or perceive the laser spot on their wings during the laser-spot heating experiments (**Supplementary Fig. 43**).

**Basking experiment**

In the experiment (**Fig. 7**), a dark-adapted adult butterfly was placed in an arena with a thermally transparent window. The temperature of the arena was controlled by a Peltier thermoelectric cooler. A Xenon lamp, which simulates sunlight, illuminated the substrate of the arena uniformly from above with an angle of ~15° to the vertical.

All the species used in this experiment perform "lateral basking"[24,25]: when the lamp was turned on, the butterflies closed their wings and tilted sideways to expose their thoraces and ventral wing surfaces to the illumination. The beginning of a lateral basking event was scored when the wings were tilted to be approximately perpendicular to the direction of illumination.

The cessation of basking was scored when the butterflies restored the vertical orientation of the wings. The basking events were recorded simultaneously by a visible camera and a thermal camera. The wing and thoracic temperatures were extracted directly from the thermal camera readings at the wing veins and thoraces, respectively, taking advantage of the fact that both the thoraces and wing veins have thermal emissivity approaching unity.

We have complied with all relevant ethical regulations for animal testing and research.

**Acknowledgements** Jaret Daniels, Patrick Griffith, Masaru K. Hojo, Tracy Magellan, Douglas Mullins, Tiago Quental, Jinhua Tan, and Xiujing Yu helped collect the insects analyzed in this research. We thank Mark Cornwall, Rod Eastwood, Thomas Dai, and James Crall for advice, Mary Salcedo for encouraging us to study hemolymph flow in butterfly wings, and Joshua Sanes for suggesting that we use methylene blue to stain neurons in butterfly wings. Our research was supported by the National Science Foundation (no. PHY-1411445 awarded to NY and NEP, no. DEB-0447242 awarded to NEP), and the Air Force Office of Scientific Research (no. FA9550-14-1-0389 through the Multidisciplinary University Research Initiative program and no. FA9550-16-1-0322 through the Defense University Research Instrumentation Program awarded to NY). RAC was supported by the Graduate Research Fellowship Program (GRFP) of the National Science Foundation. Measurements were carried out in part at the Center for Functional Nanomaterials, Brookhaven National Laboratory, which is supported by the U.S. Department of Energy, Office of Basic Energy Sciences, under contract DE-SC0012704.

**Author Contributions** CCT, GDB, NEP, and NY conceived and designed the experiments. CCT, NNS, JNP, NEP, and NY carried out the experiments. CCT, RAC, CR, NEP, and NY developed theoretical models, conducted numerical simulations, and analyzed the data. CCT, RAC, NEP, and NY wrote the manuscript, with input from all co-authors. All authors discussed the results and commented on the manuscript.

**Competing Interests**

The authors declare no competing interests.

**Data Availability**

The source data underlying Figs. 1, 2, 4, 6, and 7 and Supplementary Figs. 4, 6-9, 32, 33, 35, 37-39, 42, and 43 are provided as a Source Data file.

**Figure Legends**

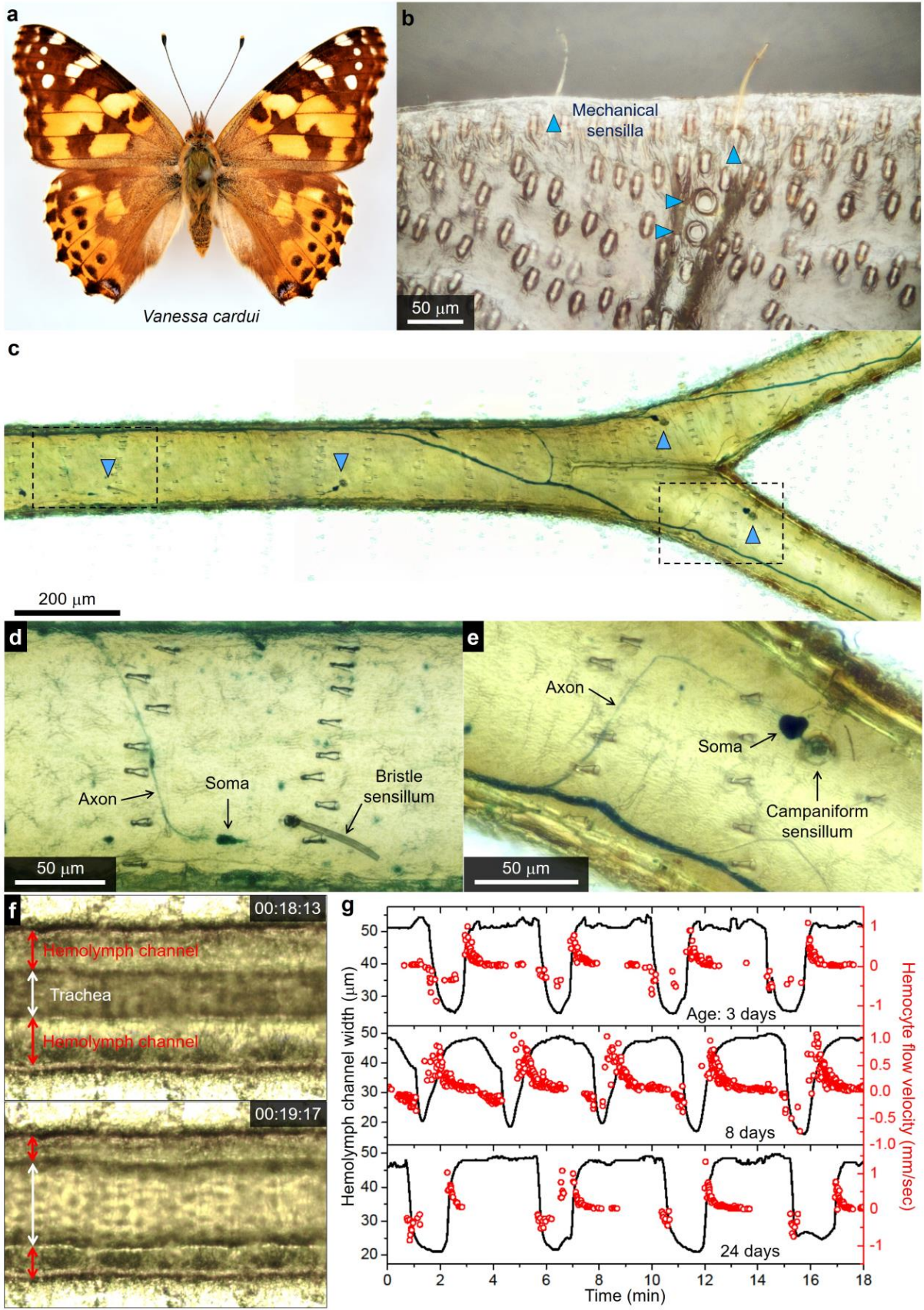

**Figure 1 | Living components of butterfly wings. a** Photo of a male painted lady butterfly *Vanessa cardui*. **b** Photo showing mechanical sensilla (indicated by blue arrows) near the wing margin of an adult *V. cardui*, including two disc-shaped campaniform sensilla and two hair-shaped bristle sensilla. **c** Photo showing nerve fibers and mechanical sensors, stained with methylene blue, in the cubital veins of the forewing of an adult *V. cardui*. Regions in the dashed frames are enlarged in **d** and **e** to show, respectively, one bristle sensillum and one campaniform sensillum as well as associated somata (or cell bodies) and axons. **f** Two snapshots of a video to show variations of the width of the hemolymph channels in the medial vein of the right forewing of an adult *V. cardui*. **g** Observed variations of hemolymph channel width (black solid curves) and instantaneous hemocyte flow velocity (red circles) as a function of time for individual *V. cardui* butterflies at three different ages (3, 8, and 24 days after eclosion). Positive sign of flow velocity indicates movement of hemocytes from the base toward the tip of the wing; negative sign indicates movement of hemocytes from the tip toward the base of the wing. **Supplementary Movie 3** shows tidal flow of hemolymph in a radial vein of *V. cardui*. Source data are provided as a Source Data file.

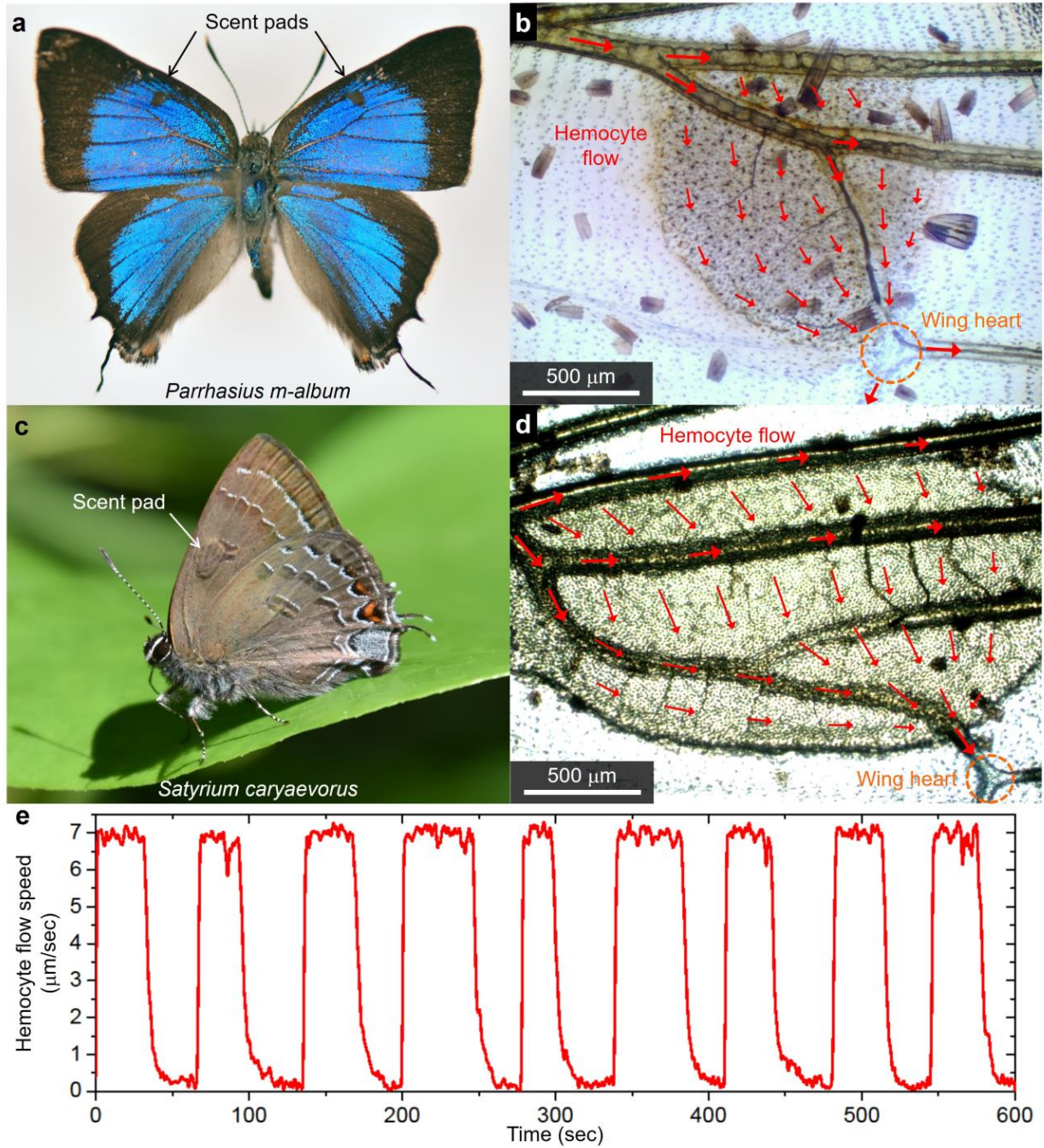

**Figure 2 | Hemolymph flow in scent pads. a** Photo of a male white-M hairstreak, *Parrhasius m-album*. **b** Photo showing the scent pad of *P. m-album*, an associated "wing heart", and observed pattern of hemocyte flow in the scent pad. **c** Photo of a male hickory hairstreak, *Satyrium caryaevorus*. **d** Photo showing the scent pad of *S. caryaevorus*, an associated "wing heart", and observed pattern of hemocyte flow in the scent pad. **e** Observed average hemocyte flow speed as a function of time in the scent pad of *S. caryaevorus*. **Supplementary Movies 4** and **5** show the beating of the wing heart and movement of hemocytes within the scent pad of *P. m-album*. **Supplementary Movies 6** and **7** show the movement of hemocytes within the scent pad of *S. caryaevorus*. Source data are provided as a Source Data file.

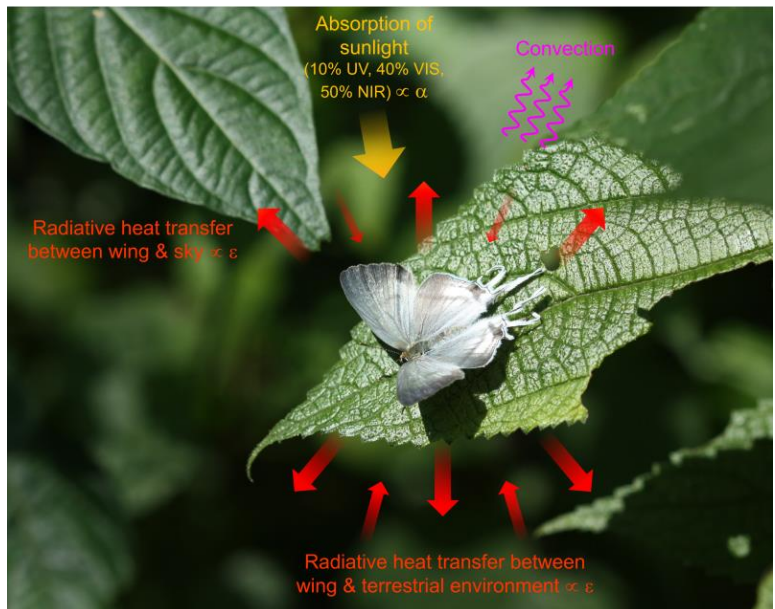

**Figure 3 | Thermodynamics of butterfly wings.** A female *Hypolycaena hatita* basking in the sun. Steady-state wing temperature is the result of a balance between absorption of sunlight (orange arrow), convective heat loss to the surrounding air (purple arrows), and radiative energy exchange between the heated wing and the relatively cold surrounding environment (red arrows). Heating by sunlight is determined by solar absorptivity, and ability to dissipate heat through thermal radiation is governed by thermal emissivity; therefore, understanding the properties of butterfly wings in the solar spectrum (where the ultraviolet, visible, and near-infrared components contain ~10%, ~40%, and ~50% of the solar energy, respectively) and in the thermal radiation spectrum ($\lambda = 3 - 20$ μm) is crucial to understand the thermodynamics of butterfly wings. Dino J. Martins photographed the *Hypolycaena hatita* shown in this figure.

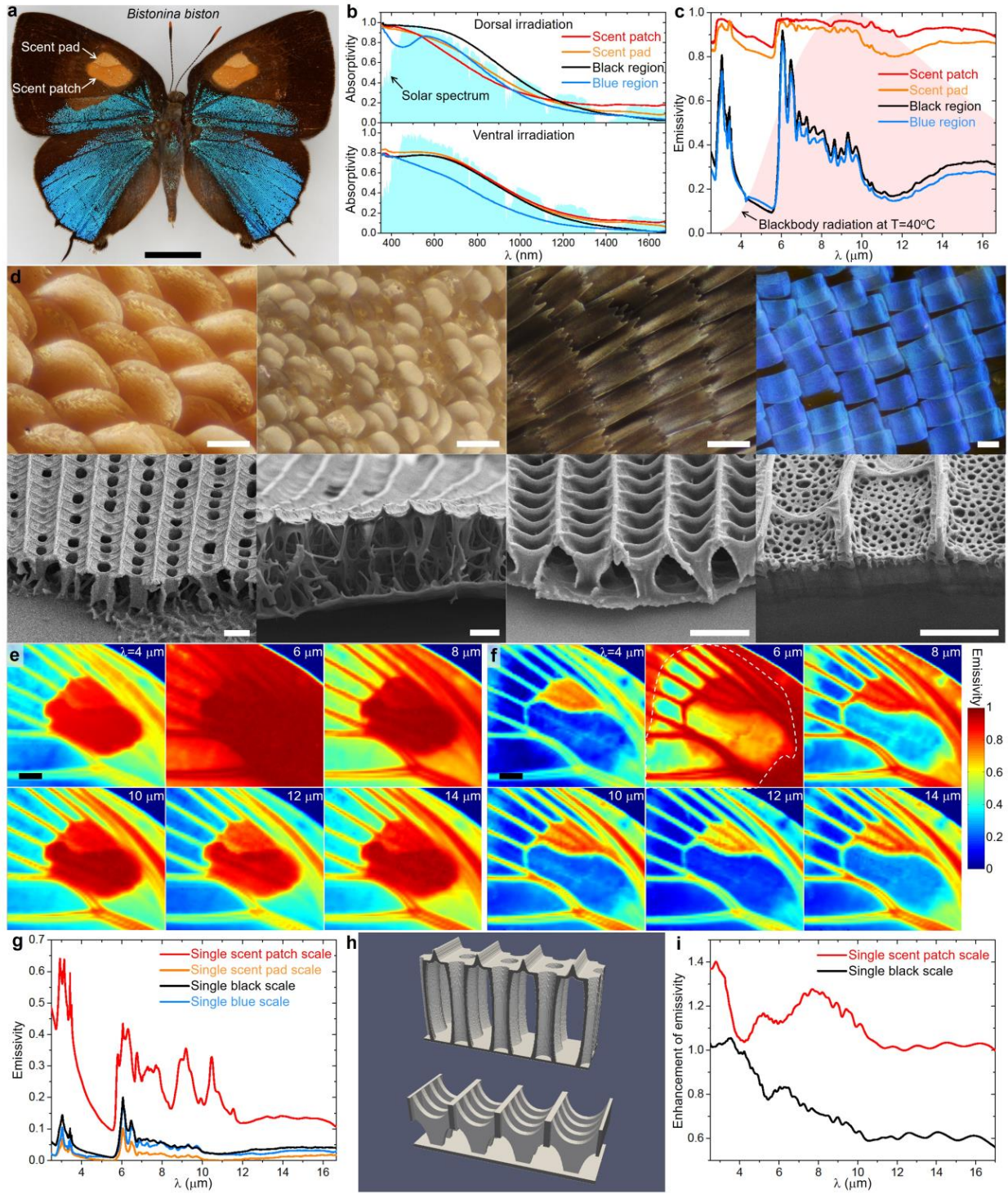

**Figure 4 | Highly heterogeneous distribution of solar absorptivity and thermal emissivity on butterfly wings. a** Photo of a male *Bistonina biston*. Each forewing has a pheromone-producing androconial organ consisting of a scent pad and a scent patch. Scalar bar: 5 mm. **b** Solar absorption spectra measured from different regions of the *B. biston* forewing. Normalized solar spectrum is also plotted (light blue shaded region). **c** Thermal emissivity spectra measured from different regions of the *B. biston* forewing. Normalized thermal radiation spectrum of a blackbody at

T=40ºC is also plotted (light red shaded region). **d** Optical images (top row) and SEM images (bottom row) of four types of wing scales on the forewing of a male *B. biston*. From left to right are scent patch, scent pad, black, and blue scales. Scalar bars in the top/bottom row: 50 μm/2 μm. **e** Distribution of thermal emissivity over the butterfly wing at a few mid-infrared wavelengths, showing that the scent patch, scent pad, and veins are highly emissive over the entire thermal radiation spectrum. **f** When wing scales have been removed (from the area indicated by the dashed curve), high emissivity of the scent patch disappears, while high emissivity of the scent pad and veins remains. Scalar bars in **e** and **f**: 1 mm. **g** Measured thermal emissivity spectra of individual scales from different regions of the wing. **h** Models of the scent patch (top) and black (bottom) scales, according to SEM images in **d**, used in full-wave simulations. **i** Simulated enhancement of thermal emissivity for individual scent patch and black scales. Spectra shown in **b**, **c**, and **g** are average of results measured from multiple locations on the specimen. Andrei Sourakov photographed the *Bistonina biston* shown in **a**. Source data are provided as a Source Data file.

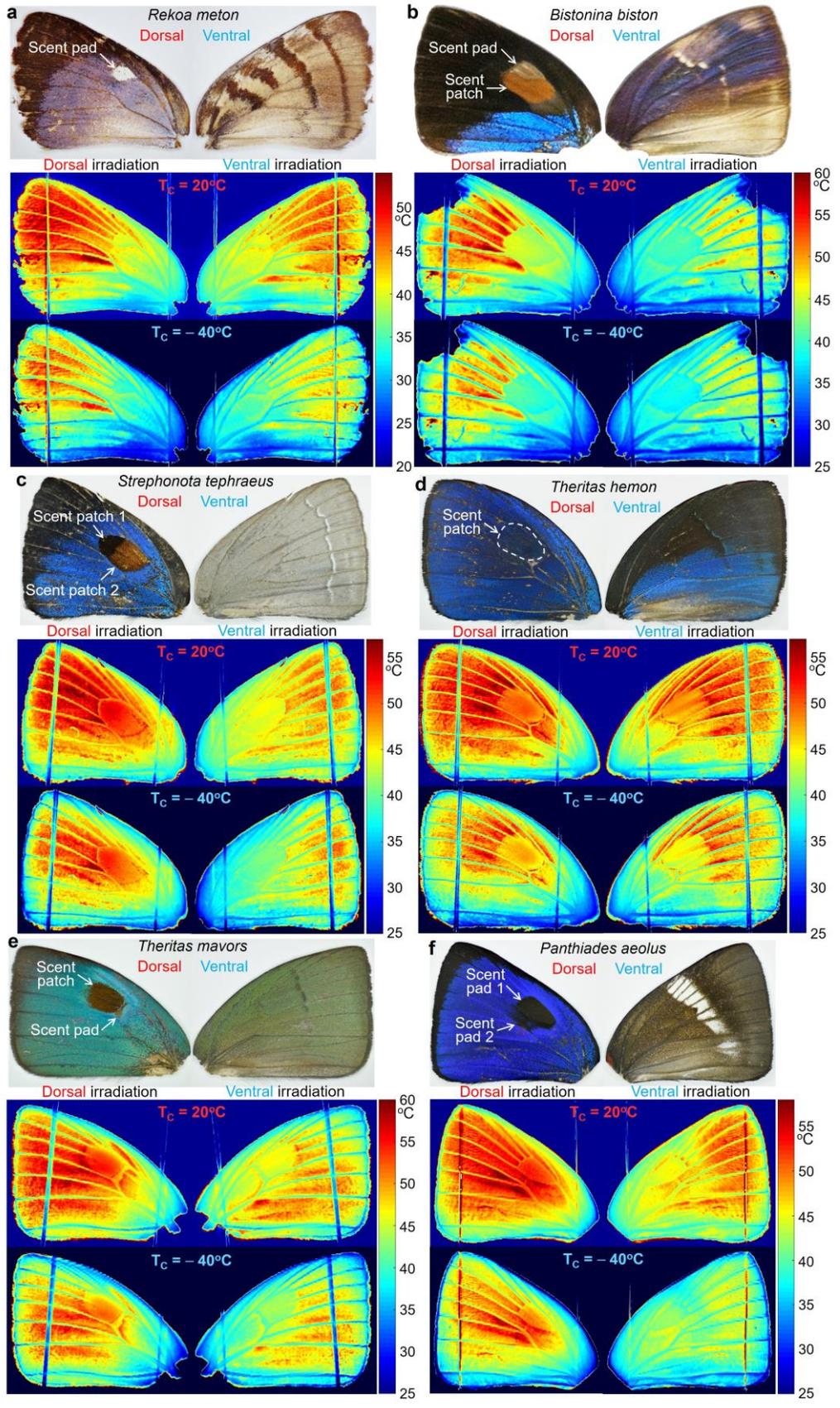

**Figure 5 | Temperature distributions on the forewing of six species of Eumaeini butterflies**. The temperature distributions were obtained by processing infrared flux images taken by a thermal camera with measured two-dimensional emissivity, transmissivity and reflectivity maps of the butterfly wings (**Supplementary Fig. 36**). The thermodynamic steady state was reached within 10 seconds of the exposure of the wings to the illumination due to their small thermal capacity (**Supplementary Fig. 39**). Within each sub-figure, the top shows photos of the dorsal and ventral forewing, and the bottom shows wing temperature distributions under four experimental conditions (i.e., light simulating full sun incident on the dorsal or ventral wing surface, and cryostat temperature, $T_c$, of +20°C or −40°C). $T_c$ = +20°C simulates an environmental condition with a relatively high ambient temperature and humidity; $T_c$ = −40°C simulates an environmental condition with a relatively low ambient temperature and humidity. The exact combinations of ambient temperature and humidity simulated by $T_c$ = +20°C or −40°C are provided in **Supplementary Fig. 35**. One individual per species has been used in this study.

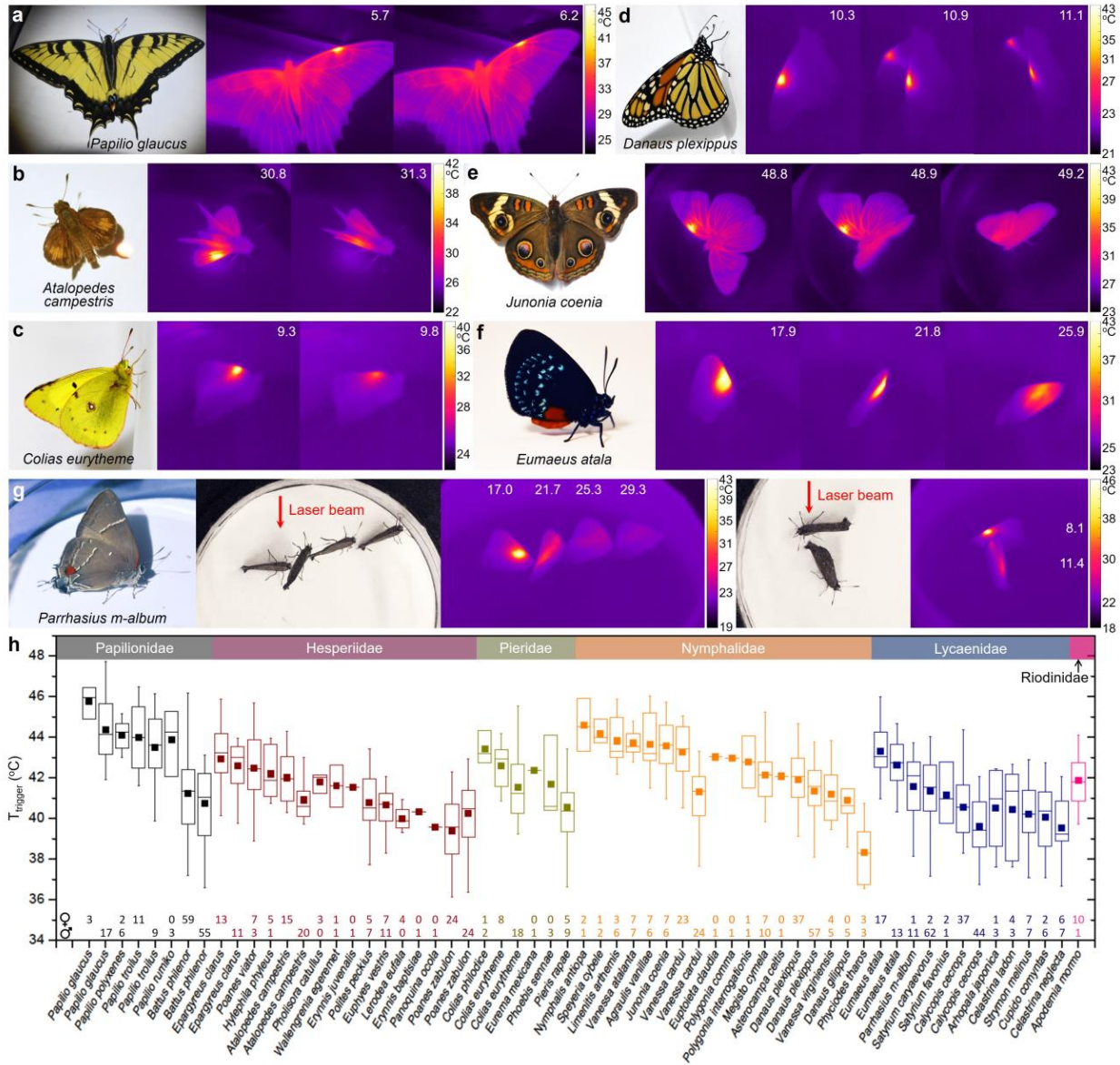

**Figure 6 | Displacement response induced by local heating of butterfly wings. a-g** Photos and thermal camera images showing the butterflies and their typical displacement responses when the wings are locally heated by a laser beam. For example, adults of *Eumaeus atala* tend to tilt their wings to reduce light intensity and thus temperature on the wing; those of *Parrhasius m-album* either walk away from the laser beam path or change their body orientation to reduce wing temperature. Numbers in the thermal camera images indicate time in seconds. **Supplementary Movies 8** and **9** show a number of species of butterflies during the experiments recorded using the thermal camera. **h** Box plot of temperatures triggering the response, $T_{trigger}$, for 862 individuals representing 50 species in six families of butterflies (Papilionidae: 5 species, Hesperiidae: 13 species, Pieridae: 5 species, Nymphalidae: 16 species, Lycaenidae: 10 species, Riodinidae: 1 species); the number of individuals tested per sex and per species is indicated along the bottom horizontal axis. $T_{trigger}$ for each individual was calculated by using the first ten responses for each

individual to estimate a line of best fit, from which we extracted the intercept at $t = 0$. We determined this to be the baseline $T_{trigger}$ for each individual, the average of these being the baseline $T_{trigger}$ for each species. When the total number of tested individuals for a given species was >20, $T_{trigger}$ for males and females was calculated and reported separately. In the box plot, the lower and upper boundaries of the box indicate the 25[th] and 75[th] percentiles, a line within the box marks the median, and a square within the box marks the mean. Whiskers below and above the box indicate the 1[st] and 99[th] percentiles. Source data are provided as a Source Data file. Source data are provided as a Source Data file.

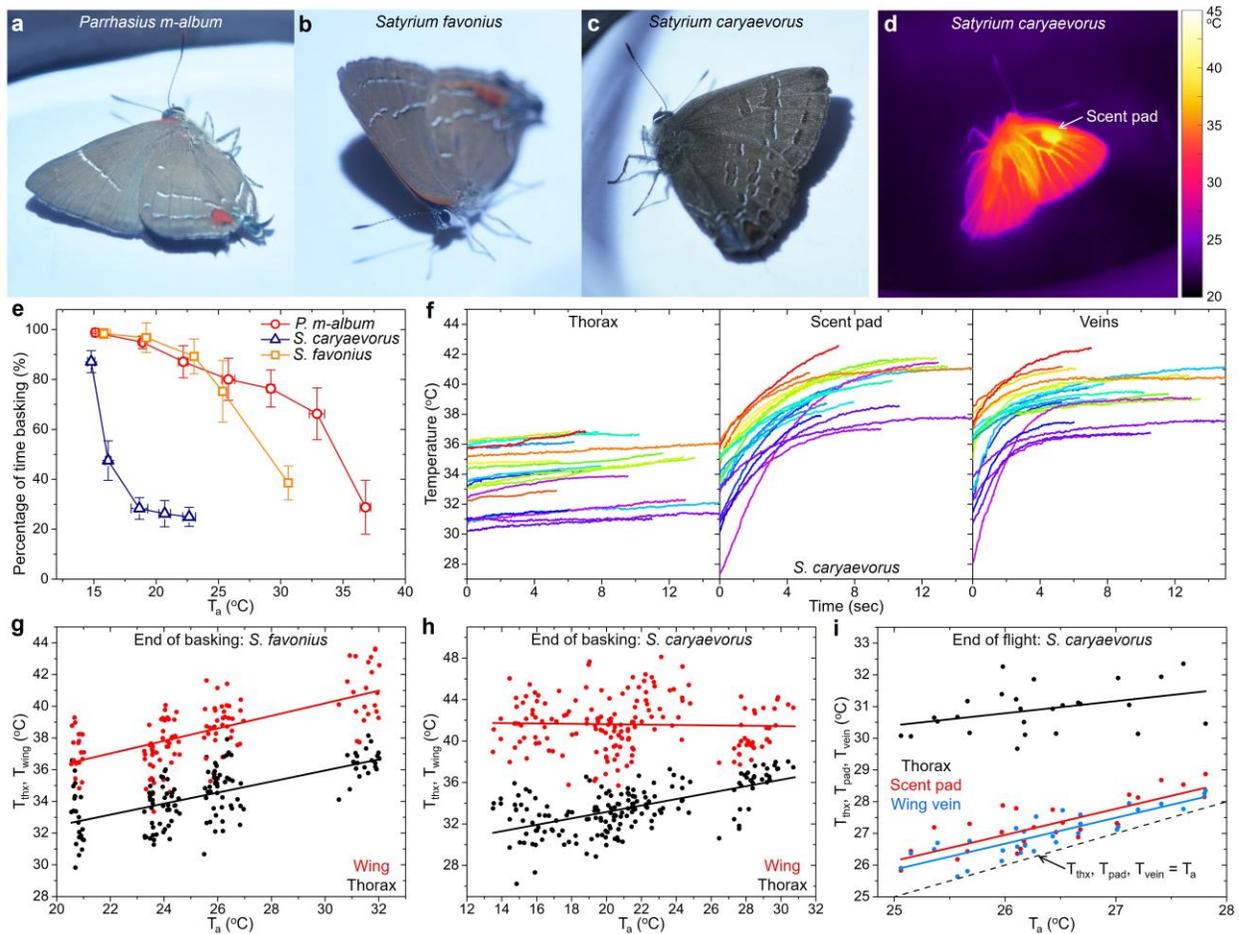

**Figure 7 | Basking behavior of three species of hairstreak butterflies. a-c** Photos showing lateral basking of *P. m-album, S. favonius*, and *S. caryaevorus*. **d** Thermal camera image of an adult of *S. caryaevorus* (uncorrected for the emissivity distribution on the butterfly) at the end of a representative basking event, showing that the scent pads and veins are much hotter than the thorax. **e** Summary of the basking experiments showing that the percentage of time spent basking decreases as the ambient temperature, $T_a$, increases for the three species of hairstreak butterflies, and that under the same conditions the *S. caryaevorus* butterflies ($n = 6$) require far less basking than do those of *P. m-album* ($n = 10$) and *S. favonius* ($n = 4$). **Supplementary Movies 10** and **11** show, respectively, basking of *S. caryaevorus* and *P. m-album* at low and high ambient

temperatures. Error bars represent one standard deviation of uncertainty. **f** Recorded temperature increases as a function of time during 20 basking events for an individual of *S. caryaevorus*. The experimental conditions are $T_a = 21.5^\circ$C and lamp power of 66.0 mW/cm$^2$ (corresponding to 2/3 of full sun). **g** and **h** Temperatures of the wing and thorax at the cessation of basking for *S. favonius* (n=4) and *S. caryaevorus* (n=6), respectively, with lamp intensity corresponding to 2/3 of full sun and for a variety of ambient temperatures. **i** Data showing the temperatures of the thorax, scent pad, and wing veins at the point when a *S. caryaevorus* butterfly stops flying. Source data are provided as a Source Data file.